\documentclass[12pt]{article}
\usepackage{graphicx}
\usepackage[all,2cell]{xy}          
\usepackage{amsthm}
\usepackage{tikz-cd}
\usepackage{mathtools}
\usepackage{mathrsfs}
\usepackage{enumerate}
\usepackage[colorlinks=true]{hyperref}
\hypersetup{allcolors=[rgb]{0.1,0.1,0.4}}

\usepackage{txfonts}

\usepackage{makeidx}
\makeindex

\usepackage[all,2cell]{xy}
\UseAllTwocells
\xyoption{arc}
\xyoption{rotate}

\usepackage{hyperref}   
\hypersetup{colorlinks, 
breaklinks,             
linkcolor=black,        
urlcolor=black}         

\newcommand{\fcat}[1]{\mathbf{#1}}      

\newcommand{\N}{\mathbb{N}}           

\renewcommand{\r}{\mathbb{R}}

\newcommand{\Set}{\fcat{Set}}           

\newcommand{\ladj}{\dashv}              

\newcommand{\oppair}[4]{%
\xymatrix{%
#1 \ar@<.5ex>[r]^-{#3} &{#2}\ar@<.5ex>[l]^-{#4}%
}}

\newcommand{\oppairi}[4]{%
\xymatrix@1{%
#1 \ar@<.5ex>[r]^{#3} &{#2}\ar@<.5ex>[l]^{#4}%
}}

\newcommand{\parpair}[4]{%
\xymatrix{%
#1 \ar@<.5ex>[r]^{#3} \ar@<-.5ex>[r]_{#4} &#2%
}}

\newcommand{\parpairi}[4]{%
\xymatrix@1{%
#1 \ar@<.5ex>[r]^{#3} \ar@<-.5ex>[r]_{#4} &#2%
}}

\newcommand{\adjn}[4]{%
\xymatrix{
#1 \ar@{}[d]|\ladj \ar@<1ex>[d]^{#4} \\
#2 \ar@<1ex>[u]^{#3}
}}

\newcommand{\hadjnli}[4]{%
\xymatrix@1{
#1 \ar@<1.1ex>[r]^-{#3} \ar@{}[r]|-\bot &#2 \ar@<1.1ex>[l]^-{#4}}}

\newcommand{\hadjnri}[4]{%
\xymatrix@1{
#1 \ar@<1.1ex>[r]^-{#3} \ar@{}[r]|-\top &#2 \ar@<1.1ex>[l]^-{#4}}}

\newtheorem{exem}{Exemple}

\newtheorem{defn}{D\'efinition}

\newtheorem{definition}{D\'efinition}

\renewcommand{\1}{{\bf 1}}

\newcommand{\mc}{\mathscr{C}}

\newcommand{\ds}{\displaystyle}

\def\subclassname{{\bfseries Mathematics Subject Classification
(2000)}\enspace}
\def\subclass#1{\par\addvspace\medskipamount{\rightskip=0pt plus1cm
\def\and{\ifhmode\unskip\nobreak\fi\ $\cdot$
}\noindent\subclassname\ignorespaces#1\par}}
\usepackage[utf8x]{inputenc}
\newcommand{\bv}{\begin{verbatim}}
\newcommand{\env}{\end{verbatim}}

\usepackage{listings}
\begin{document}
\title{Accessible bridge between category theory and functional programming}

\author{ Fethi Kadhi\footnote{\'Ecole Nationale des Sciences de l'Informatique, 	Manouba University.}} 
\date{}
\maketitle
\begin{abstract}
La programmation monadique pose une difficulté énorme pour un grand nombre de programmeurs.
\`A la lumière de la théorie des catégories, on présente une nouvelle vision de l'utilisation des monades en programmation fonctionnelle.
Cette vision est clarifiée par un bon nombre d'exemples  cod\'es en Haskell.
\end{abstract}
\tableofcontents
\section{Introduction}
L'application de la théorie des catégories aux sciences de l'informatique a donn\'e naissance \`a la programmation fonctionnelle.
De nos jours les programmeurs utilisent, parfois sans être conscients, des idée et des stratégies dont l'origine est l'approche fournie par la théorie des catégories. Cependant, on remarque l'existence de deux publics presque disjoints. Un public qui connaît bien les fondements de la théorie des catégories sans connaître son lien aux sciences de l'informatique. Un autre public très conscient de l'importance de la théorie des catégorie en tant que outil d'inspiration et d'innovation mais il trouve une difficulté énorme pour accéder aux fondements théoriques des outils qu'il utilise. Le but de ce travail est de fournir un pont accessible des deux cot\'es,
c'est \`a dire accessible par les deux publics.
Haskell se trouve \`a la tête des langages de programmation purement fonctionnelle. Le pont, ainsi construit, est riche en exemples écrits avec le code Haskell.
\section{Les langages de programmation}\label{intro}
\subsection{Programmation impérative}
En informatique, la programmation impérative est un paradigme de programmation qui décrit les opérations en séquences d'instructions exécutées par l'ordinateur pour modifier l'état du programme. Ce type de programmation est le plus répandu parmi l'ensemble des langages de programmation existants. Il se différencie de la programmation déclarative (dont la programmation logique ou encore la programmation fonctionnelle sont des sous-ensembles).
La quasi-totalité des processeurs qui équipent les ordinateurs sont de nature impérative : ils sont faits pour exécuter une suite d'instructions élémentaires, codées sous forme d'opcodes (pour operation codes). L'ensemble des opcodes forme le langage machine spécifique à l'architecture du processeur. L'état du programme à un instant donné est défini par le contenu de la mémoire centrale à cet instant.

Les langages de plus haut niveau utilisent des variables et des opérations plus complexes, mais suivent le même paradigme. Puisque les idées de base de la programmation impérative sont à la fois conceptuellement familières et directement intégrées dans l'architecture des microprocesseurs, la grande majorité des langages de programmation est impérative.\\
La plupart des langages de haut niveau comporte cinq types d'instructions principales :
\begin{enumerate}
	\item la séquence d'instructions
\item l'assignation ou affectation
\item l'instruction conditionnelle
\item la boucle
\item les branchements
	\end{enumerate}
Exemples de langages impératifs: Fortran, Algol, Cobol, Basic, Pascal, C, Ada, Smaltalk, C++ et Objective, Perl, Tcl, Python, PHP, Java, JavaScript
\subsection{Programmation fonctionnelle}
La programmation fonctionnelle est un paradigme de programmation de type déclaratif qui considère le calcul en tant qu'évaluation de fonctions mathématiques.
Comme le changement d'état et la mutation des données ne peuvent pas être représentés par des évaluations de fonctions la programmation fonctionnelle ne les admet pas, au contraire elle met en avant l'application des fonctions, contrairement au modèle de programmation impérative qui met en avant les changements d'état.

Le paradigme fonctionnel n'utilise pas de machine à états pour décrire un programme, mais une composition de fonctions qui agissent comme des  ``boîtes noires''  que l'on peut imbriquer les unes dans les autres. Chaque boîte possédant plusieurs paramètres en entrée mais une seule sortie, elle ne peut sortir qu'une seule valeur possible pour chaque n-uplet de valeurs présentées en entrée. Ainsi, les fonctions n'introduisent pas d'effets de bord. Un programme est donc une application, au sens mathématique, qui ne donne qu'un seul résultat pour chaque ensemble de valeurs en entrée. Un avantage important des fonctions sans effet de bord est la facilité de les tester unitairement. Par ailleurs, l'usage généralisé d'une gestion de mémoire automatique simplifie la tâche du programmeur.

Les langages fonctionnels emploient des types et des structures de données de haut niveau comme les listes extensibles. Il est ainsi généralement possible de réaliser facilement des opérations comme la concaténation de listes, ou l'application d'une fonction à une liste — le parcours de la liste se faisant de façon récursive — en une seule ligne de code.

Un mécanisme puissant des langages fonctionnels est l'usage des fonctions d'ordre supérieur. Une fonction est dite d'ordre supérieur lorsqu'elle peut prendre des fonctions comme arguments (aussi appelées callback) ou renvoyer une fonction comme résultat. On dit aussi que les fonctions sont des objets de première classe, ce qui signifie qu'elles sont manipulables aussi simplement que les types de base. Les programmes ou fonctions qui manipulent des fonctions correspondent, en mathématiques, aux fonctionnelles. Les opérations de dérivation et d'intégration en sont deux exemples simples. Les fonctions d'ordre supérieur ont été étudiées par Alonzo Church et Stephen Kleene dans les années 1930, à partir du formalisme du lambda-calcul, qui a influencé la conception de plusieurs langages fonctionnels, notamment celle de Haskell. La théorie des catégories cartésiennes fermées constitue une autre approche pour caractériser les fonctions, à l'aide de la notion de problème universel. Le langage Anubis s'appuie sur cette théorie.
 Alors que l'origine de la programmation fonctionnelle peut être trouvée dans le lambda-calcul, le langage fonctionnel le plus ancien est Lisp, créé en 1958 par McCarthy. Lisp a donné naissance à des variantes telles que Scheme (1975) et Common Lisp (1984) qui, comme Lisp, ne sont pas ou peu typées. Des langages fonctionnels plus récents tels ML (1973), Haskell (1990), OCaml, Erlang, Clean et Oz, CDuce, Scala (2003), F\# ou PureScript (2013), Agda sont fortement typés.

\subsection{Haskell}
Son nom vient du mathématicien et logicien Haskell Curry. Il a été créé en 1990 par un comité de chercheurs en théorie des langages intéressés par les langages fonctionnels et l'évaluation paresseuse. Le dernier standard est Haskell 2010 : c'est une version minimale et portable du langage conçue à des fins pédagogiques et pratiques, dans un souci d'interopérabilité entre les implémentations du langage et comme base de futures extensions. Le langage continue d'évoluer en 2020, principalement avec GHC, constituant ainsi un standard de facto comprenant de nombreuses extensions.
Haskell est un langage de programmation purement fonctionnel, 
ce qui signifie que, de façon inhérente, aucun effet de bord n'est autorisé, ni les entrées/sorties, ni les affectations de variables, ni les exceptions. La plupart des langages fonctionnels encouragent ce style, mais Haskell l'impose dans tout code qui ne signale pas explicitement par son type qu'il admet des effets de bord et qui grâce à ce mécanisme en prévient et en circonscrit les dangers.
\section{La catégorie Hask}
Dans ce paragraphe, nous allons explorer le langage Haskell \`a la lumière de la théorie des catégories.
Haskell est une catégorie Hask. Les objets de Hask sont les types de Haskell. Les flèches de Hask sont les fonctions de Haskell.
Les questions qui se posent alors sont
\begin{enumerate}
	\item Quels sont les types de Haskell?
	\item Comment construire des fonctions entre les types de Haskell?
	\item Comment coder en Haskell la composition de deux fonctions?
	\item Comment construire un endo-foncteur dans Hask qui satisfait les axiomes d'un foncteur en théorie des catégories?
\end{enumerate}
 \subsection{Les types de Haskell}
Haskell est un langage fonctionnel et il est strictement typé, ce qui signifie que le type de donnée utilisé dans l'ensemble de l'application sera connu du compilateur au moment de la compilation.
\begin{enumerate}
	\item {\bf Int} est une classe de types représentant les données de types Integer. Chaque nombre entier compris entre $2^{31}-1=2147483647$ et $-2147483647$ appartient à la classe de type Int. Int utilise une mémoire de longueur $32$ bits dont une case est réservée pour le signe.
	\item {\bf Integer} peut être considéré comme un sur-ensemble de Int. Cette valeur n'est limitée par aucun nombre, donc un entier peut être de n'importe quelle longueur sans aucune limitation.
	\item {\bf Float}
	\item {\bf Double} est un nombre à virgule flottante avec deux chiffres après la virgule.
	\item {\bf Bool} est un type booléen. Il peut être True ou False. Exécutez le code suivant pour comprendre comment le type Bool fonctionne dans Haskell:
   %
	\item {\bf Char} représente les caractères. Tout ce qui se trouve dans un guillemet simple est considéré comme un caractère.
	\item {\bf EQ} La classe de type EQ est une interface qui fournit la fonctionnalité pour tester l'égalité d'une expression. Toute classe Type qui souhaite vérifier l'égalité d'une expression doit faire partie de cette classe Type EQ.

Toutes les classes de type standard mentionnées ci-dessus font partie de cette classe d'égalisation. Chaque fois que nous vérifions une égalité à l'aide de l'un des types mentionnés ci-dessus, nous effectuons en fait un appel à la classe de type EQ.
\item {\bf Ord} est une autre classe d'interface qui nous donne la fonctionnalité de commande. Tous les types que nous avons utilisés jusqu'à présent font partie de cette interface Ord. Comme l'interface EQ, l'interface Ord peut être appelée en utilisant $">", "<", "<=", ">="$, "compare".
\item {\bf Show} a une fonctionnalité pour imprimer son argument sous forme de chaîne. Quel que soit son argument, il affiche toujours le résultat sous forme de chaîne. 
\item {\bf Read} fait la même chose que Show, mais elle n'imprimera pas le résultat au format String. 
\item {\bf Enum}  est un autre type de classe Type qui active la fonctionnalité séquentielle ou ordonnée dans Haskell. Cette classe Type est accessible par des commandes telles que Succ, Pred, Bool, Char, etc.
\item{\bf Integral} peut être considéré comme une sous-classe de la classe de type Num. La classe Num Type contient tous les types de nombres, tandis que la classe de type Integral n'est utilisée que pour les nombres entiers. Int et Integer sont les types de cette classe Type.

\item {\bf Flottant}
Comme Integral, Floating fait également partie de la classe Num Type, mais il ne contient que des nombres à virgule flottante. Par conséquent, Float et Double relèvent de cette classe de type.
\end{enumerate}
\subsection{Les fonctions de Haskell}
\begin{lstlisting}[language=haskell]
-- exemple 1
add :: Integer -> Integer -> Integer   --function declaration 
add x y =  x + y                       --function definition 
-- x,y sont de type Integer, le resultat est de type Integer
main = do 
   putStrLn "The addition of the two numbers is:"  
   print(add 2 5)    --calling a function 
   print(add 12 5)    --calling a function
	
-- Exemple 2
fact :: Int -> Int 
fact 0 = 1 
fact n = n * fact ( n - 1 ) 

main = do 
   putStrLn "The factorial of 5 is:" 
   print (fact(5))
	
-- Exemple 3
roots :: (Float, Float, Float) -> (Float, Float)  
roots (a,b,c) = (x1, x2) where 
   x1 = e + sqrt d / (2 * a) 
   x2 = e - sqrt d / (2 * a) 
   d = b * b - 4 * a * c  
   e = - b / (2 * a)  
main = do 
   putStrLn "The roots of our Polynomial equation are:" 
   print (roots(1,-8,6))
   print (roots(1,-2,1))

-- Exemple 5
-- Lambda Expression
main = do 
   putStrLn "The successor of 4 is:"  
   print ((\x -> x + 1) 4)
-- Head function and others
main = do 
   let x = [1..10]
   let y=[10..20]   
   putStrLn "Our list is:"  
   print (x) 
--putStrLn "The first element of the list is:" 
   print (head x)
   print (last x)
   print (reverse x)
   print (length x)
   print (maximum x)
   print (sum x)
   print (product x)
   putStrLn "Does it contain 786?" 
   print (elem 786 (x))
   print(x++y)-- l'operateur ++ pour concatener deux listes en Haskell
\end{lstlisting}

\subsection{Loi de composition en Hask}
\begin{verbatim}
g :: Int -> Bool 
f :: Bool -> String --fog est bien defini
-- On declarer les foncctions puis donner la definition de chaque fonction
g x = if x `rem` 2 == 0 
   then True 
   else False 
f x = if x == True 
   then "This is an even Number" 
else "This is an ODD number" 

main = do 
   putStrLn "Example of Haskell Function composition" 
   print ((f.g)(16))
   print ((f.g)(27))
\end{verbatim}

\subsection{La flèche identité en Hask}
La flèche identité $id$ est prédéfinie en Haskell. Dans l'exemple précédent nous avons composé deux fonctions $f$ et $g$.
On peut vérifier que la composition avec $id$ ne change pas le résultat de la fonction.
\begin{verbatim}
-- Flèche identité
g :: Int -> Bool 

g x = if x `rem` 2 == 0 
   then True 
   else False 
main = do 
   putStrLn "Example of Haskell Function composition" 
   print(g(16))
   print ((id.g)(16))
   print ((g)(27))
   print ((id.g)(27))

=========================
g :: Int -> Bool 
f :: Bool -> String --fog est bien defini

g x = if x `rem` 2 == 0 
   then True 
   else False 
f x = if x == True 
   then "This is an even Number" 
else "This is an ODD number" 

main = do 
   putStrLn "Example of Haskell Function composition" 
   print ((f.g.id)(16))
   print ((id.f.g)(27))
\end{verbatim}
\subsection{Foncteurs en Hask}
Les foncteurs sont introduits dans la théorie des catégories et leur concept est utilisé en Haskell (ainsi que dans de nombreuses branches des mathématiques). Un foncteur dans la théorie des catégories est un opérateur. Cet opérateur renvoie les objets d'une catégorie source vers une catégorie de destination, et renvoie également les flèches de la catégorie source à la catégorie de destination. Un foncteur en Haskell est un tel opérateur. En Haskell, les objets sont des types et les flèches sont des fonctions d'un type à un autre. Dans Haskell, les catégories source et destination contiennent des types. En outre, les catégories source et destination peuvent être considérées comme étant les mêmes.

Ainsi, si j'ai un type $a$, un type $b$ et une flèche (c'est-à-dire une fonction) $p:a -> b$, alors un Functor $f$ effectue les applications suivantes:
\begin{enumerate}[i]
	\item  $f$ fait correspondre le type $f a$ au type $a$.
\item $f$ fait correspondre le type $f b$ au type $b$.
\item $f$ fait correspondre la fonction $fp:fa ->fb$ à la fonction $p: a -> b$.
\end{enumerate}

Veuillez noter que les objets $a$ et $b$ peuvent être de même type ou de types différents. Notez également que, $f$ représente le Foncteur et non une fonction. Les fonctions ici sont désignées par $->$.

Un Foncteur en Haskell est représenté par un constructeur de type armé d'un mécanisme de mappage de fonction. Le constructeur de type fait le mappage des types (points i et ii dans le paragraphe précédent) et le mécanisme de mappage de fonction fait le mappage des fonctions (point iii dans le paragraphe précédent).
Voici un programme en Haskell qui crée et utilise un foncteur qui s'appelle MyF
\begin{exem}\item
\begin{verbatim}
module Main where
 
data MyFunctor a = MyF a deriving (Show)
 
instance Functor MyFunctor where
   fmap f (MyF x) = MyF (f x)
 
main :: IO ()
main  =
   do
      putStrLn "Program begins."
 
      let thing1 = MyF 45
      print thing1
      print (fmap (*2) thing1)
      print (fmap (+1) thing1)
      print (fmap (:[]) thing1)
      print (fmap (\x -> 2*x+1:[]) thing1)
      print thing1
 
      putStrLn "Program ends."
\end{verbatim}

Here is the output of this program:

\begin{verbatim}
Program begins.
MyF 45
MyF 90
MyF 46
MyF [45]
MyF [91]
MyF 45
Program ends.
\end{verbatim}
\end{exem}
In the previous program, the data MyFunctor statement declares/creates a type constructor. This type constructor is called MyFunctor.
 As a type constructor, MyFunctor maps a type $a$ to the type MyFunctor $a$. In this example, MyFunctor $a$ is always equal to MyF $a$. Also, as a type constructor, MyFunctor maps a type $b$ to the type MyFunctor $b$. In this example, $MyFunctor b$ is always equal to $MyF b$.

So what is missing is a way to map a function $a -> b$ to $MyFunctor a -> MyFunctor b$. This remaining ingredient is provided by  the $fmap$ function.

Now, Haskell already knows about the $fmap$ function. Haskell already has built-in the declaration of this function:
\begin{verbatim}
class Functor f where
   fmap :: (a -> b) -> f a -> f b
\end{verbatim}

or equivalently:
\begin{verbatim}
fmap :: Functor f => (a -> b) -> f a -> f b
\end{verbatim}

In the previous program, we implement the function $fmap$ as follows:
\begin{verbatim}
fmap f (MyF x) = MyF (f x)
\end{verbatim}
Here $f$ denotes a function $a -> b$ ($f$ does not denote a Functor here!). 
 So, what we are saying to Haskell here is that we want to map functions $a -> b$ in such a way so that they operate on the value after MyF. Thus, a function f :: a->b that operates on a MyFunctor a type, will give as output a MyFunctor b type such that the function $f$ will operate on the type that exists on the right of MyF.

The correspondence here is the following:
\begin{enumerate}
	\item  The function $(a -> b)$ in the declaration corresponds to the function $f$ in our definition.
	\item The Functor $f$ in the declaration corresponds to the data type constructor value MyF in our definition.
	\item The type $a$ in the declaration corresponds to the type $x$ in our definition.
	\item The type $b$ in the declaration corresponds to the type $f x$ in our definition.
	\item The type $f a$ in the declaration corresponds to the type $MyF x$ in our definition.
	\item The type $f b$ in the declaration corresponds to the type $MyF (f x)$ in our definition.
	\item From the declaration, we see that $fmap$ takes two arguments, a function $(a -> b)$ and an object of type $f a$ and returns an object of type of $f b$ as output. 
	\item From our definition, we correspondingly see that indeed $fmap$ takes two arguments, a function $f$ and an object of type $MyF x$ and returns an object of type of $MyF (f x)$ as output.
\end{enumerate}

Let us study the examples that are given for the use of MyFunctor in the main program.

First of all, we create $thing1$ which is equal to $MyF 45$ and thus is of type MyFunctor.
 Here we made use of the type constructor in order to map 45 (which is an Integer) to MyF 45 which is an object of type MyFunctor (and, especially, MyF). So, we have a mapping of type Int to MyF Int.

Then, we print $thing1$, just to see that everything is OK. This is possible because we used the deriving (Show) sub-statement in the declaration of MyFunctor.

Then, we print the result of fmap (*2) thing1, which is the result of applying the function (*2) to thing 1. thing1 is of type MyFunctor and (*2) is a function $Int -> Int$. We can rewrite $(*2)$ as $(\textbackslash x -> x*2)$.
We have already told fmap how to function in such a case:
\begin{verbatim}
fmap f (MyF x) = MyF (f x)
\end{verbatim}

$fmap$ is to take our (*2) function and apply it to thing1 such that MyF 45 will become MyF ((*2) 45), thus MyF 90.

The same holds true for the next statement, where we apply the function (+1) to thing1. Again, the definition for fmap that we have provided will give the result MyF 46.

Now, the next statement is a little bit more interesting. It prints the result of fmap (:[]) thing1, which is the result of applying the function :[] to thing1. This function has different types for its input and output. :[] is a function of Int -> [Int], thus from Int to a list of Ints. But as in the previous two statements, fmap knows to apply :[] to the Int after MyF. Thus the result is MyF [45].

The last statement prints the result of fmap $(\backslash x -> 2*x+1:[])$ thing1, thus of applying a function that does all the previous operations to thing1. The function is of type $Int -> [Int]$ and the result is $MyF [91]$.

At the end of the program, we print thing1 again, in order to show that it has not changed. We have immutability in Haskell and, anyway, thing1 was only used as input to the $fmap$ function during the course of the program.

Dans l'exemple suivant on va vérifier que les deux axiomes d'un foncteur sont satisfaits.
\begin{exem}
\item 
\begin{verbatim}
module Main where
 
data MyFunctor a = F1 a
                 | F2 [a]
                 | F3 (a,a)
                 | F4 a
                 deriving (Show)
 
instance Functor MyFunctor where
   fmap f (F1 x)  = F1 (f x)
   fmap f (F2 xs) = F2 (map f xs)
   fmap f (F3 x) = F3 (f (fst x), f (snd x))
   fmap f (F4 x) = F1 (f x)
 
main :: IO ()
main  =
   do
      putStrLn "Program begins."
 
      let thing1 = F1 10
      print thing1
      print (fmap (*2) thing1)
      print thing1
 
      let thing2 = F2 [100,1000,10000,100000]
      print thing2
      print (fmap (*2) thing2)
      print thing2
 
      let thing3 = F3 (400,500)
      print thing3
      print (fmap (*2) thing3)
      print thing3
 
      let thing4 = F4 200
      print thing4
      print (fmap (*2) thing4)
      print (fmap (id) thing4)-- le premier axiome d'un foncteur
      print (fmap ((*2).(*3)) thing4)--fmap(f0g)
      print (fmap (*2) (fmap (*3) thing4))--fmap(f)0fmap(g)
      print thing4
 
      putStrLn "Program ends."
\end{verbatim}
Ce programme donne le output suivant:
\begin{verbatim}
Program begins.
F1 10
F1 20
F1 10
F2 [100,1000,10000,100000]
F2 [200,2000,20000,200000]
F2 [100,1000,10000,100000]
F3 (400,500)
F3 (800,1000)
F3 (400,500)
F4 200
F1 400
F1 200
F1 1200
F1 1200
F4 200
Program ends.
\end{verbatim}
\end{exem}
\section{Monade en Haskell}
\subsection{Monades}
On ne peut pas prétendre la connaissance de la programmation fonctionnelle sans maîtriser cette notion de monade.
\begin{defn}
Une monade sur une catégorie $\mc$ consiste en un triplet $(T,\eta,\mu)$ avec
$T:\mc\longrightarrow\mc$ est un endo-foncteur (son domaine est le même que son co-domaine).
 $\eta:1_\mc\Longrightarrow T$ et $\mu:T^2\Longrightarrow T$ sont deux transformations naturelles
 telles que les deux diagrammes suivants commutent:
$$
\xymatrix{
T \ar@{=>}[r]^{\eta T} \ar@{=>}[dr]_{1_T}&T^2\ar@{<=}[r]^{T\eta } & T\\
    & T \ar@{<=}[u]_{\mu}\ar@{<=}[ur]_{1_T} &
} \hspace{3cm}
\xymatrix{
T^3 \ar@{=>}[r]^{\mu T} \ar@{=>}[d]_{T\mu} & T^2 \ar@{=>}[d]^{\mu}      \\
T^2 \ar@{=>}[r]_{\mu}      &  T
} 
$$
C'est \`a dire $\mu\circ\eta T=\mu\circ T\eta =1_T$ et $\mu\circ\mu T=\mu\circ T\mu $.
$\eta$ s'appelle l’unité de la monade et $\mu$ s'appelle sa multiplication.
\end{defn}
\begin{exem}\label{powerset}
Soit $P$ le foncteur qui associe \`a un ensemble $X$ l'ensemble de ses parties.
$$\begin{array}{lrll}
P:&\Set&\longrightarrow&\Set\\
  &X&\mapsto&P(X)\\
	&f\downarrow&         &\downarrow P(f)\\
	&Y&\mapsto&P(Y)
\end{array}$$
Si $A$ est une partie de $X$ alors $P(f)(X)=\{f(x):x\in X\}=f(X)$.\\
Soient $\eta:1_\Set\Longrightarrow P$ et $\mu:P^2\Longrightarrow P$ deux transformations naturelles de composantes
suivant un ensemble $X$:
$$\begin{array}{lrll}
 \eta_X:&X&\longrightarrow&P(X)\\
  &x&\mapsto&\{x\}
	\end{array}
	\hspace{1cm}\mbox{ et } 	\hspace{1cm}
	\begin{array}{lrll}
 \mu_X:&P(P(X))&\longrightarrow&P(X)\\
  &\{A_1,\dots,A_n\}&\mapsto&\cup_i A_i
	\end{array}
	$$
	Soit $A$ une partie d'un ensemble $X$. D'une part, on a $(1_{P})_X(A)=1_{PX}(A)=A$. D'autre part,
	\begin{eqnarray*}
	(\mu\circ P\eta)_X(A)&=&(\mu_X\circ P(\eta_X))(A)\\
	                      &=&\mu_X(\eta_X(A))\\
												&=&\mu_X(\{\{a\}:a\in A\}\\
												&=&\cup_{a\in A}\{a\}\\
												&=& A\\
												&=&(1_P)_X(A)
	\label{eq:}
	\end{eqnarray*}
	Il s'ensuit que $\mu\circ P\eta=1_P$.
	
	Le même raisonnement permet de voir que 
	 le triple $(P,\{.\},\cup)$ est une monade sur $\Set$. On remarque dans cet exemple que la monade $P$ permet l'extension d'une fonctions du $X\stackrel{f}{\rightarrow}Y$ en une fonction $PX\stackrel{Pf}{\rightarrow}PY$ entre des ensembles "plus grands".
\end{exem}
\subsection{Monades en Haskell}
In Haskell a monad is represented as:
\begin{enumerate}
	\item  a type constructor (call it m).
	\item a function called ``return'' that builds values of that type $(a -> m a)$.
	\item a function that combines values of that type with computations that produce values of that type to produce a new computation for values of that type $(m a -> (a -> m b) -> m b)$. This function is known as "bind" and is written "$>>=$".
	\end{enumerate}
	Ce qui s’écrit en Haskell:
	\begin{verbatim}
	class Monad m where
  (>>=)  :: m a -> (  a -> m b) -> m b
  (>>)   :: m a ->  m b         -> m b
  return ::   a                 -> m a
	\end{verbatim}
	To be a proper monad, the return and $(>>=)$ functions must work together according to three laws:
\begin{enumerate}
	\item $(return\: x) >>= f ==== f x$
	\item $m >>= return ==== m$
	\item $(m >>= f) >>= g ==== m >>= (\backslash x -> f x >>= g)$
\end{enumerate}
The first law requires that return is a left-identity with respect to $>>=$. The second law requires that return is a right-identity with respect to $>>=$. The third law is a kind of associativity law for $>>=$. Obeying the three laws ensures that the semantics of the do-notation using the monad will be consistent.
Any type constructor with return and bind operators that satisfy the three monad laws is a monad. In Haskell, the compiler does not check that the laws hold for every instance of the Monad class. It is up to the programmer to ensure that any Monad instance they create satisfies the monad laws.
Ce qui s’écrit en Haskell:
\begin{verbatim}
	return a >>= k                  =  k a
  m        >>= return             =  m
   m        >>= (\x -> k x >>= h)  =  (m >>= k) >>= h
	\end{verbatim}
	De façon plus simple:
	Dans Haskell, une monade $m$ est un type conteneur. Parmi toutes les foncions qui peuvent s'appliquer \`a $m$, on distingue la fonction
$return$ qui consiste \`a injecter un objet de type $a$ dans le conteneur $m$ pour obtenir un objet de type $ma$. Ça ressemble un peu \`a l'unit\'e:
$$\begin{array}{lrll}
 \eta_X:&X&\longrightarrow&P(X)\\
  &x&\mapsto&\{x\}
	\end{array}$$
	\begin{verbatim}
	return  ::  ( Monad  m )  =>  a  ->  m  a
	\end{verbatim}
	Si j'ai une pomme $a$, je peux la mettre dans une boîte $m$ pour obtenir $ma$.\\
	Si $ma$ est contenu \`a son tour dans $m$ alors on obtient $m(ma)$. La fonction $join$ consiste \`a combiner les objets de $ma$ pour obtenir un objet de type $ma$, 
\begin{verbatim}
join  ::  ( Monad  m )  =>  m  ( m  a )  ->  m  a
\end{verbatim}	
		ce qui ressemble un peu \`a la multiplication de la monade $P$ définie par:
$$\begin{array}{lrll}
 \mu_X:&P(P(X))&\longrightarrow&P(X)\\
  &\{A_1,\dots,A_n\}&\mapsto&\cup_i A_i
	\end{array}
	$$	
	Si les pommes sont stockées en cageots dans le frigo alors on peut oublier les cageots
	pour les considérer  stockées directement dans le frigo.\\
	Ce que fait $bind$ est de prendre un conteneur de type $ma$ et une fonction de type $(a -> m b)$ . Il applique d'abord $m$ \`a la fonction (ce qui donnerait un $(ma -> m(m b)$ ), puis applique une jointure au résultat pour obtenir un conteneur de type $(mb)$ . Son type et sa définition en Haskell est alors
	\begin{verbatim}
	(>>=) :: (Monad m) => m a -> (a -> m b) -> m b
	\end{verbatim}
	\subsection{Monade List $\left[\:\right]$ }
\begin{exem}\item
\begin{verbatim}
main = do
   print([1..10] >>= return)
\end{verbatim}

Ce code Haskell a pour output:
	\begin{verbatim}
     [1,2,3,4,5,6,7,8,9,10]
\end{verbatim}
\end{exem}
Dans cet exemple, on vérifie que la première loi d'une monade est satisfaite pour le foncteur .
En effet, l'ordre de la composition ici est: on applique ``return'' aux contenu de la liste puis on applique $>>=$ \`a la liste obtenue
ce qui correspond \`a $\mu\circ P(\eta)=1_P$.	
\begin{exem}\item
\begin{verbatim}
main = do
   print([1..10] >>= (\x -> if odd x then [x*2] else []))
\end{verbatim}
\end{exem}
Ce code Haskell a pour output:
	\begin{verbatim}
     [2,6,10,14,18]
\end{verbatim}
En fait, Haskell construit une liste $\left[\:\right]$ suivant sa définition d'une monade, c'est pour cette raison 
$\left[\:\right]$ se trouve dans le domaine de définition de sa fonction ``bind'' codée par $(>>=)$.
Dans cet exemple, $\left[\:\right]$ joue le rôle de $m$.
$a$ est la liste des integers de 1 \`a 10 et $b$ est la liste des 
integers $2,6,10,14,18$. La taille de la liste est flexible avec le changement de la taille des données causé par l'application 
  de la fonction $(>>=)$ after la fonction
	\begin{verbatim}
 (\x -> if odd x then [x*2] else [])
\end{verbatim}
Si on veut voir exactement l'action de la fonction $(>>=)$ alors une première idée est de la supprimer du code précédent pour visualiser
l'\'etat de la liste juste sous l'action de la fonction [x*2].
\begin{verbatim}
 (\x -> if odd x then [x*2] else [])
\end{verbatim}
Malheureusement, le code obtenu retourne une erreur:
\begin{verbatim}
Couldn't match expected type: (a1 -> [a1]) -> a0
                  with actual type: [a2]
\end{verbatim}
Un simple astuce permet de neutraliser l'effet de $(>>=)$. Il suffit de composer par la fonction return:
\begin{verbatim}
main = do
   print([1..10] >>=return. (\x -> if odd x then [x*2] else []))
\end{verbatim}

Ce code Haskell a pour output:
	\begin{verbatim}
     [[2],[],[6],[],[10],[],[14],[],[18],[]]
\end{verbatim}
Le même output peut être obtenu en regardant une $\left[\:\right]$ comme un foncteur. L'action de la fonction
\begin{verbatim}
 (\x -> if odd x then [x*2] else [])
\end{verbatim}
peut être codée par $fmap$:
\begin{verbatim}
main = do
   print(fmap (\x -> if odd x then [x*2] else []) [1..10])
\end{verbatim}
Ainsi, il est clair que la fonction $(>>=)$ transforme la liste des listes objets:
\begin{verbatim}
     [[2],[],[6],[],[10],[],[14],[],[18],[]]
\end{verbatim}
en une liste d'objets:
\begin{verbatim}
     [2,6,10,14,18]
\end{verbatim}
Un autre exemple pour visualiser encore l'action de la fonction bind $(>>=)$. 
On va appliquer, \`a la liste $\left[\1..10\right]$  d'abord la fonction:
\begin{verbatim}
    \x -> [x,x+1]
\end{verbatim}
On peut utiliser $fmap$:
\begin{verbatim}
main = do
   print(fmap (\x -> [x,x+1]) [1..10])
	\end{verbatim}
	Ce code a pour output cette liste de listes:
	\begin{verbatim}
    [[1,2],[2,3],[3,4],[4,5],[5,6],[6,7],[7,8],[8,9],[9,10],[10,11]]
\end{verbatim}
La fonction bind $(>>=)$ est construite de sorte quelle transforme la liste des listes d'objets en une liste d'objets:
\begin{verbatim}
main = do
   print([1..10] >>=(\x -> [x,x+1]))
\end{verbatim}
Ce code a pour output:
\begin{verbatim}
    [1,2,2,3,3,4,4,5,5,6,6,7,7,8,8,9,9,10,10,11]
\end{verbatim}
Si on veut supprimer les objets qui se répètent alors on peut appliquer \`a la liste obtenue la fonction $nub$:
 \begin{verbatim}
    import Data.List-- pour importer la fonction nub
main = do
   print \$ nub ([1..10] >>=(\x -> [x,x+1]))
\end{verbatim}
Ce code a pour output:
\begin{verbatim}
    [1,2,3,4,5,6,7,8,9,10,11]
\end{verbatim}
\subsection{Application}
Pour un entier $n$, on veut déterminer touts les triplets $(x,y,z)$ qui satisfont la relation de Pythagore
$z^2=x^2+y^2$ avec $z\leq n$.\\
Soit $\phi:\N\to\N\times\N\times\N$, avec
$$\phi(n)=\{(x,y,z):z\leq n \texttt{ et } z^2=x^2+y^2\}$$
On peut voir $\phi$ comme
$$\ds\phi(n)=\cup_{1\leq x\leq n}(\cup_{1\leq y\leq n}(\cup_{1\leq z\leq n}\{(x,y,z):z^2=x^2+y^2\}))$$
La seule fonction Haskell qui permet de fusionner cette réunion de réunions est la fonction bind codée en Haskell par $(>>=)$.
Ainsi, on peut construire la fonction $\phi$ en Haskell de la façon suivante:
\begin{verbatim}
phi :: Integer -> [(Integer, Integer, Integer)]
phi n =
  [1 .. n] >>= (\x ->  [1 .. n] >>= (\y ->  [1 .. n] >>= (\z ->
      if x^2 + y^2 == z^2 then [(x,y,z)] else [])))
main = do
print(phi 9)
\end{verbatim}
Ce programme a pour output:
\begin{verbatim}
[(3,4,5),(4,3,5)]
\end{verbatim}
On peut améliorer ce programme en évitant les triplets de la forme $(y,x,z)$ lorsque le triplet
$(x,y,z)$ existe déjà et en remarquant que la somme de deux carr\'es d'un même entier ne peut pas être un carr\'e:
\begin{verbatim}
 pythagoreanTriples :: Integer -> [(Integer, Integer, Integer)]
pythagoreanTriples n =
  [1 .. n] >>= (\x ->  [x+1 .. n] >>= (\y ->  [y+1 .. n] >>= (\z ->
      if x^2 + y^2 == z^2 then  (x,y,z) else [])))
main = print \$ pythagoreanTriples 25
\end{verbatim}
Ce programme a pour output:
\begin{verbatim}
 [Running] runhaskell "c:\haskellexamples\tempCodeRunnerFile.hs"
[(3,4,5),(5,12,13),(6,8,10),(7,24,25),(8,15,17),
(9,12,15),(12,16,20),(15,20,25)]

[Done] exited with code=0 in 0.383 seconds
\end{verbatim}
\subsection{Monade Maybe}
\begin{definition}
Une donnée $y$  est de type (Maybe a) si et seulement si $y$ s’écrit sous la forme:
$y=\mbox{ Just}(x)$ avec $x$ est de type a ou $y=\mbox{ Nothing}$

\begin{verbatim}
data Maybe a = Nothing | Just a
              deriving (Eq, Ord)
instance (Eq a) => Eq (Maybe a)
Nothing == Nothing = True
Nothing == Just y = False
Just x == Nothing = False
Just x == Just y = (x == y)
instance (Ord a) => Ord (Maybe a)
Nothing <= Nothing = True
Nothing <= Just y = True
Just x <= Nothing = False
Just x <= Just y = (x <= y)							
\end{verbatim}
\end{definition}
La monade (Maybe) permet d’étendre un type a en un type (Maybe a) qui contient les objets de a augment\'e d'un objet qui s'appelle Nothing.
En Haskell, une monade est un foncteur muni de deux transformations naturelles: une unit\'e $return$ 
et une fonction bind  $>>=$. La question est donc c'est quoi $return$ et c'est quoi $>>=$ dans le cas de Maybe?\\
Voici la réponse de Haskell \`a cette question:
\begin{verbatim}
-- La fonction return
return :: a -> Maybe a
return x  = Just x
-- La fonction bind
(>>=) :: Maybe a -> (a -> Maybe b) -> Maybe b
(>>=) m g = case m of
                   Nothing -> Nothing
                   Just x  -> g x
\end{verbatim}

Ceci veut donc dire que si le premier état correspondait à l’état d’erreur Nothing, alors la fonction de transition d’état $a -> Maybe b$ ne sera jamais exécutée et on renverra Nothing. Ceci fournit une abstraction de la gestion d’erreur très utile afin de réduire les occurrences d’écriture d’instructions redondantes par le développeur.

Par exemple la fonction $f: x\longmapsto\frac{1}{x}$ est définie sur $\r^*$. Si on associe \`a $0$ l'objet Nothing alors $f$ est définie sur $\r$. On dit que $f$ est une fonction totale. Considérons le code Haskell suivant:
\begin{verbatim}
f ::  Float -> Maybe Float
f x = if x >= 0 then Just(1-sqrt(x)) else Nothing

main = do 
      print \$ (f(3))
      print \$ (f(-3)) 
      print \$ (f(0))
      print \$ (f(-1))
\end{verbatim}
Ce programme a pour output
\begin{verbatim}
Just (-0.7320508)
Nothing
Just 1.0
Nothing
\end{verbatim}
Remarquer que Just est une instance d'un foncteur. On pourra alors construire avec $fmap$ une fonction
$g: Maybe\: Float\to Maybe\: Float$:
 \begin{verbatim}
g :: Maybe Float -> Maybe Float
g x = fmap (\x -> log x ) (x)
main = do 
      print $ (g(Just(5)))
      print $ (g(Just(0))) 
      print $ (g(Nothing))
      print $ (g(Just(-1)))
\end{verbatim}
Ce programme a pour output
\begin{verbatim}
Just 1.609438
Just (-Infinity)
Nothing
Just NaN
\end{verbatim}
Remarquons que $g$ est composable avec la fonction $f$ puisque le domine de $f$ est le codomaine de $g$.
\begin{verbatim}
f ::  Float -> Maybe Float
f x = if x >= 0 then Just(1-sqrt(x)) else Nothing
g :: Maybe Float -> Maybe Float
g x = fmap (\x -> log x ) (x)
main = do 
      print $ ((g.f)(3))
      print $ ((g.f)(-3)) 
      print $ ((g.f)(0))
      print $ ((g.f)(1))
\end{verbatim} 
Ce code a pour output
\begin{verbatim}
Just NaN
Nothing
Just 0.0
Just (-Infinity)
\end{verbatim}
On remarque que $g$ retourne des différents résultats pour les cas dégénérés: Just NaN, Nothing, Infinity. 
La monade (Maybe) permet de simplifier les calculs en Just(x) ou Nothing:
\begin{verbatim}
f ::  Float -> Maybe Float
f x = if x >= 0 then Just(1-sqrt(x)) else Nothing
g :: Maybe Float -> Maybe Float
g mx = case mx of 
      Nothing -> Nothing
      Just(x) -> if x > 0 then Just(log(x)) else Nothing
main = do 
      print $ ((g.f)(3))
      print $ ((g.f)(-3)) 
      print $ ((g.f)(0))
      print $ ((g.f)(1))
\end{verbatim}
Ce code a pour output
\begin{verbatim}
Nothing
Nothing
Just 0.0
Nothing
\end{verbatim}
Dans une composition de plusieurs fonctions, il suffit que l'une des fonctions retourne Nothing pour que toute la composition retourne Nothing. On dit que la monade Maybe propage l'erreur.\\
Si on veut extraire la valeur $x$ de Just(x) alors on peut composer avec la fonction $h$ qui renvoie le type
Maybe Float vers le type Float:
\begin{verbatim}

f ::  Float -> Maybe Float
f x = if x >= 0 then Just(sqrt(x)) else Nothing

g :: Maybe Float -> Maybe Float
g mx = case mx of 
      Nothing -> Nothing
      Just(x) -> if x > 0 then Just(log(x)) else Nothing
			
h::Maybe Float->Float
h mx = case mx of
    Nothing -> 0
    Just(x) -> x
main = do 
      print $ ((h.g.f)(0.5))
      print $ ((h.g.f)(-3)) 
      print $ ((h.g.f)(5))
      print $ ((h.g.f)(1))
\end{verbatim}
Ce code donne le output suivant:
\begin{verbatim}
-0.34657362
0.0
0.804719
0.0
\end{verbatim}
En Haskell, la monade Maybe permet de traiter le problème de manque d'informations dans une base de données.
Supposons par exemple que notre base de données est formée d'un ensemble de listes. Nous voulons interroger cette base sur la tête de chaque liste. Chaque liste vide peut être la cause d'une erreur qui dérange l’exécution du programme.
La monade Maybe permet de résoudre ce problème une fois pour toutes. On peut redéfinir la fonction head de sorte qu'elle retourne
Nothing lorsque la liste est vide.
\begin{verbatim}
import Data.List
safehead :: [Int] -> Maybe Int
safehead [] = Nothing
safehead xs = Just (head xs)
main = do
    print $ (safehead([]))
    print $ (safehead([6,1,2]))
\end{verbatim}
Ce programme a pour output:
\begin{verbatim}
Nothing
Just 6
\end{verbatim}
Supposons maintenant que notre base de données est formée d'une listes de numéros de téléphones:
\begin{center}
\begin{tabular}{|l|l|}
\hline
Ali&96552233\\
\hline
Belgacem&98555111\\
\hline
Salha&27211211\\
\hline
Mohsen&\\
\hline
Massaoud&55222333\\
\hline
\end{tabular}
\end{center}
On veut interroger cette base de donnée sur le numéro d'un nom donn\'e. Si le nom ne figure pas dans la base alors on risque
d'avoir un message d'erreur. La monade Maybe permet de retourner toujours une réponse quitte a retourner Nothing lorsque le nom n'existe pas dans la liste. Dans Haskell, la fonction qui interroge une liste sur une telle donnée s'appelle lookup:
\begin{verbatim}
phonebook :: [(String, String)]
phonebook = [ ("Ali","96552233"),
              ("Belgacem",  "98555111"),
              ("Salha", "27211211"),
              ("Mohsen", "" ),
			  ("Massaoud","55222333")]
main = do
    print $ lookup "Ali" phonebook
    print $ lookup "Salem" phonebook
\end{verbatim}
Ce code a pour output
\begin{verbatim}
Just "96552233"
Nothing
\end{verbatim}
\end{document}